\documentclass[11pt]{article}

\usepackage[margin=1in]{geometry}
\usepackage[round,authoryear]{natbib}
\usepackage[colorlinks,citecolor=blue,urlcolor=blue,linkcolor=blue]{hyperref}
\usepackage{amsmath,amsfonts,amsthm,amssymb,mathtools}
\usepackage{booktabs,array}
\usepackage{enumitem}
\usepackage{xcolor}
\usepackage{authblk}

\newtheorem{assumption}{Assumption}[section]
\newtheorem{theorem}{Theorem}[section]
\newtheorem{remark}{Remark}[section]
\newtheorem{proposition}{Proposition}[section]
\newtheorem{lemma}{Lemma}[section]
\newtheorem{corollary}{Corollary}[section]
\newtheorem{definition}{Definition}[section]

\newcommand{\EE}{\mathbb{E}}
\newcommand{\oP}{o_P}

\newcommand{\iid}{\overset{\mathrm{i.i.d.}}{\sim}}
\newcommand{\Ltwo}{L^2(P_0)}
\newcommand{\Psitrue}{\Psi(P_0)}
\newcommand{\Psihat}{\hat{\Psi}}

\newcommand{\Dstar}{D^{*}}

\newcommand{\EICj}{\mathrm{EIC}_j}

\newcommand{\calP}{\mathcal{P}}

\newcommand{\norm}[1]{\left\|#1\right\|_{P_0}}
\newcommand{\Rtwo}{R_2}
\newcommand{\Rrem}{R_{\mathrm{rem}}}

\newcommand{\VarSand}{\widehat{\mathrm{Var}}_{\mathrm{Sand}}}
\newcommand{\VarJK}{\widehat{\mathrm{Var}}_{\mathrm{JK}}}
\newcommand{\VarBoot}{\widehat{\mathrm{Var}}_{\mathrm{Boot}}}
\newcommand{\VarHC}{\widehat{\mathrm{Var}}_{\mathrm{HC}}}
\newcommand{\sigEIF}{\sigma^2_{\mathrm{EIF}}}

\newcommand{\CIJK}{\mathrm{CI}^{\mathrm{JK}}_{1-\alpha}}

\newcommand{\QY}{\bar{Q}_Y}

\newcommand{\PsiTMLE}{\hat{\Psi}_{\mathrm{TMLE}}}

\title{Variance Inference Beyond the Sandwich for Asymptotically Linear Estimators with Second-Order Remainders}

\author[1]{Lin Li\thanks{Corresponding author: \texttt{lil071@ucsd.edu}}}
\author[2]{Pengcheng Wu\thanks{\texttt{penwu@kth.se}}}
\affil[1]{Division of Biostatistics, University of California San Diego}
\affil[2]{Department of Mechatronics, KTH Royal Institute of Technology}

\date{\today}

\begin{document}
\maketitle

\begin{abstract}
First-order semiparametric inference typically uses the same influence-function linearization to justify both centering and sandwich variance estimation. We show that these are distinct questions. In a near-boundary regime, the second-order remainder can be small enough for first-order centering yet large enough to contribute $O(n^{-1})$ sampling variation. Then the usual sandwich estimator captures only the influence-function component, and Wald intervals can undercover even when the point estimator is nearly unbiased. We derive a finite-sample variance decomposition and show that, when the cross term is asymptotically negligible, sandwich consistency is characterized by the condition $n\,\mathrm{Var}(\Rrem) \to 0$: the sandwich consistently estimates total variance in the strong-decay regime where this holds, and materially underestimates it in the near-boundary regime where $n\,\mathrm{Var}(\Rrem) \to c_R > 0$. We then study two corrections. Under a leave-one-out remainder linearization condition, the jackknife consistently estimates the total variance; under a conditional Mallows-2 bootstrap condition, the pairs bootstrap does likewise. For clustered data we derive a leading-order expression showing how intra-cluster correlation magnifies the gap. Simulations for a surrogate-assisted targeted learning estimator in stepped-wedge cluster-randomized trials illustrate the regime and show substantial coverage improvement from the jackknife.


\end{abstract}

\section{Introduction}
\label{sec:intro}

A common template in semiparametric statistics is that an estimator $\Psihat$ admits a von~Mises expansion,
\begin{equation}
  \Psihat - \Psitrue
  = \frac{1}{n}\sum_{i=1}^n \Dstar(O_i) + \Rrem,
  \label{eq:ale_intro}
\end{equation}
where $\Dstar$ is an influence function and $\Rrem$ is a second-order remainder \citep{bickel1993efficient,newey1994asymptotic}. Write $\sigEIF := \EE_{P_0}[D^*(O)^2]$ for the influence-function variance. Standard practice estimates the sampling variance of $\Psihat$ by $\sigEIF/n$, yielding the familiar sandwich or Wald approximation. This reduction is justified when the remainder is negligible for \emph{variance} purposes, not merely negligible for point estimation.

The distinction matters near the product-rate boundary. In the strong-decay regime where $n\,\mathrm{Var}(\Rrem) \to 0$, the first-order linearization is sufficient for both asymptotic normality and first-order variance estimation. But many modern estimators are naturally analyzed only up to $\Rrem = O_P(n^{-1/2})$, the product-rate boundary: the remainder is small enough for near-unbiased point estimation while still large enough to contribute $O(n^{-1})$ sampling variation in the near-boundary regime. There the sandwich estimator continues to target the influence-function component only, and confidence intervals based on it can undercover even when the bias of $\Psihat$ is negligible.

This paper studies that mismatch. Our focus is not on non-negligible \emph{bias}, which has been analyzed in other nonstandard semiparametric settings \citep{cattaneo2018kernel}, but on non-negligible \emph{remainder variance}. Related finite-sample concerns have appeared empirically for TMLE and doubly robust estimators \citep{vdl2011targeted,vermeulen2015bias}. We isolate a general mechanism behind those observations: first-order linearization can be adequate for estimating the center of the sampling distribution while still being inadequate for estimating its spread.

Our contributions are fourfold. First, we derive a finite-sample variance decomposition that separates the influence-function variance, the remainder variance, and their covariance. Second, we give a sharp characterization of when the sandwich is consistent for the total variance: roughly, one needs $n\,\mathrm{Var}(\Rrem) \to 0$ when the cross term is negligible. Third, we analyze two resampling corrections. Under a leave-one-out remainder linearization condition, the jackknife consistently estimates the total variance; under a conditional Mallows-2 assumption, so does the pairs bootstrap. Fourth, for clustered data we show analytically how intra-cluster correlation can magnify the discrepancy between sandwich and total variance.

The paper separates validity of first-order \emph{centering} from validity of first-order \emph{variance estimation}. In the strong-decay regime ($n\,\mathrm{Var}(\Rrem) \to 0$), the same first-order expansion justifies both. In the near-boundary regime, however, this equivalence can fail: the estimator can remain effectively first-order for point estimation while the usual sandwich misses an $O(n^{-1})$ contribution to the sampling variance.

Our contribution is distinct from several adjacent literatures. Bias-correction work \citep[e.g.,][]{cattaneo2018kernel} addresses settings in which the center of the sampling distribution is itself non-negligibly shifted; here the center can already be first-order correct, while the failure occurs in the spread. Higher-order expansion and Edgeworth approaches seek a finer approximation to the full sampling law; here we isolate a specific $O(n^{-1})$ variance component omitted by the sandwich rather than pursuing a general higher-order approximation. Standard asymptotic-linearity results establish the expansion~\eqref{eq:ale_intro}; here that expansion is the starting point, and the question is when it is insufficient for variance estimation and how resampling recovers the missing contribution.

We do not claim a universal verification theorem for every highly adaptive learner. Instead, we isolate the variance-failure mechanism and provide sufficient conditions under which resampling recovers the total variance. The results are therefore most relevant for semiparametric inference problems in which asymptotic linearity is available but second-order remainder variation may still be non-negligible.

\section{Setup and the Near-Boundary Regime}
\label{sec:setup}

\subsection{Asymptotically Linear Estimators and the Von Mises Expansion}

Let $\mathcal{P}$ be a nonparametric model and $\Psi: \calP \to \mathbb{R}$ a
pathwise differentiable functional with efficient influence function $\Dstar
\in L^2_0(P_0)$.  We study estimators $\Psihat$ that admit the von~Mises
expansion
\begin{equation}
  \Psihat - \Psitrue
  = \frac{1}{n}\sum_{i=1}^n \Dstar(O_i) + \Rrem,
  \label{eq:ale}
\end{equation}
where $\Rrem = \Rrem({\hat{P}}, P_0)$ is a second-order remainder satisfying
$\Rrem = O_P(n^{-1/2})$ (Assumption~\ref{ass:ale}).
For brevity, we occasionally refer to such an estimator as an
asymptotically linear estimator (ALE). Throughout, we use this
shorthand for estimators admitting the first-order von~Mises
expansion~\eqref{eq:ale}; in the near-boundary regime, this
expansion remains the starting point for inference even though
classical remainder negligibility need not hold.
Throughout, we write
\[
  \bar D_n \;:=\; n^{-1}\sum_{i=1}^n D^*(O_i)
\]
for the influence-function average, and $\eta^0 = (\eta_1^0, \eta_2^0, \ldots)$
for the true nuisance functions appearing in the standard bilinear
representation of $\Rrem$ below.  When it is useful to make dependence on
nuisances explicit we write $D^*(O_i; \eta^0)$; in the unindexed notation
$D^*(O_i)$ this dependence is understood to be at the truth.
In the strong-decay
regime, $\sqrt{n}\,\Rrem = \oP(1)$ and \eqref{eq:ale} reduces to classical
asymptotic linearity; in the near-boundary regime, $\sqrt{n}\,\Rrem = O_P(1)$
and the remainder is at the same stochastic order as the influence-function
term.  The key structural property of \eqref{eq:ale} is that $\Rrem$ is
\emph{second-order} in the nuisance estimation error under standard bilinear
structure (Definition~\ref{def:remainder}).

\begin{definition}[Standard von Mises Remainder Structure]
\label{def:remainder}
We say $\Rrem$ has \emph{standard bilinear structure} if it can be written as
\begin{equation}
  \Rrem
  = \int \bigl(\hat{\eta}_1 - \eta^0_1\bigr)
         \bigl(\hat{\eta}_2 - \eta^0_2\bigr)\,dP_0
  + \oP\bigl(n^{-1/2}\bigr),
  \label{eq:bilinear}
\end{equation}
where $\eta_1, \eta_2$ are nuisance functionals with true values
$\eta_1^0, \eta_2^0$ under $P_0$ and $\hat{\eta}_k$ are their estimators.
The SA-TMLE remainder $\Rtwo$ satisfies~\eqref{eq:bilinear} with
$\eta_1 = \QY$ and $\eta_2 = g_\Delta$.  By the Cauchy--Schwarz inequality,
$|\Rrem| \le \norm{\hat{\eta}_1 - \eta^0_1} \cdot \norm{\hat{\eta}_2 - \eta^0_2}$,
so $\norm{\hat{\eta}_1 - \eta^0_1} \cdot \norm{\hat{\eta}_2 - \eta^0_2}
= \oP(n^{-1/2})$ suffices for $\sqrt{n}\,\Rrem = \oP(1)$.
\end{definition}

\begin{definition}[Product-Rate Boundary]
\label{def:boundary}
We say the estimator operates at the \emph{product-rate boundary} if each
nuisance error is of order $n^{-1/4}$ in $\Ltwo$:
\begin{equation}
  \norm{\hat{\eta}_k - \eta^0_k} = O_P(n^{-1/4}), \quad k=1,2.
  \label{eq:boundary}
\end{equation}
At the boundary the product $\norm{\hat{\eta}_1 - \eta^0_1} \cdot
\norm{\hat{\eta}_2 - \eta^0_2}$ is only $O_P(n^{-1/2})$, so the usual
sufficient condition $\oP(n^{-1/2})$ for asymptotic linearity need not hold.
The boundary is typical when
nuisance functions are estimated via parametric models in dimension growing
with $n$, or via nonparametric ensemble methods that have not yet entered
their fast-rate regime.
\end{definition}

\subsection{Regularity Conditions}

We work under the following conditions throughout Sections~\ref{sec:decomp}--\ref{sec:sand}.

\begin{assumption}[i.i.d.\ Data]
\label{ass:iid}
$O_1, \ldots, O_n \iid P_0$ with $O_i \in \mathcal{O}$.
\end{assumption}

\begin{assumption}[Von Mises Expansion]
\label{ass:ale}
The estimator $\Psihat$ admits the expansion \eqref{eq:ale} with
$\Dstar \in L^2_0(P_0)$, $\sigEIF = \EE_{P_0}[(\Dstar)^2] \in (0,\infty)$,
and $\Rrem = O_P(n^{-1/2})$.
\end{assumption}

\begin{remark}[Relation to classical asymptotic linearity]
\label{rem:ale_relation}
Classical asymptotic linearity requires $\sqrt{n}\,\Rrem = \oP(1)$, which
corresponds to the strong-decay regime of Assumption~\ref{ass:rate}
($n\cdot r_n^2 \to 0$, implying $\sqrt{n}\,\Rrem\to 0$ in probability).
In the near-boundary regime ($n\cdot r_n^2 \to c_R > 0$),
$\sqrt{n}\,\Rrem$ is allowed to remain $O_P(1)$, and therefore need not be
negligible for variance purposes; the expansion~\eqref{eq:ale}
holds with a remainder at the same stochastic order as the influence-function
term, and the classical asymptotic linearity conclusion is not guaranteed
by the first-order expansion alone.
The paper develops the inference theory for this boundary regime.
Assumption~\ref{ass:ale} imposes only $\Rrem = O_P(n^{-1/2})$, which is
compatible with both regimes.
\end{remark}

\begin{assumption}[Bounded Influence Function]
\label{ass:bounded}
$\|\Dstar\|_\infty \le M < \infty$ almost surely.
\end{assumption}

\begin{assumption}[Remainder Rate]
\label{ass:rate}
The remainder satisfies $\Rrem = r_n \cdot \xi_n$ where $r_n$ is a
deterministic rate and $\xi_n = O_P(1)$.  We say $\Rrem$ is at the
\emph{strong decay regime} if $n \cdot r_n^2 \to 0$, and at the
\emph{near-boundary regime} if $n \cdot r_n^2 \to c_R > 0$ for some constant.
\end{assumption}

\section{Finite-Sample Variance Decomposition}
\label{sec:decomp}

\subsection{The Decomposition Theorem}

\begin{theorem}[Finite-Sample Variance Decomposition]
\label{thm:decomp}
Under Assumptions~\ref{ass:iid}--\ref{ass:rate}, the sampling variance of
$\Psihat$ decomposes as:
\begin{equation}
  \mathrm{Var}\bigl(\Psihat - \Psitrue\bigr)
  \;=\;
  \underbrace{\frac{\sigEIF}{n}}_{\substack{\text{influence-function}\\\text{variance}}}
  \;+\;
  \underbrace{\mathrm{Var}(\Rrem)}_{\substack{\text{remainder}\\\text{variance}}}
  \;+\;
  \underbrace{2\,\mathrm{Cov}\!\left(\frac{1}{n}\sum_{i=1}^n \Dstar(O_i),\;\Rrem\right)}_{\text{cross term}}.
  \label{eq:decomp}
\end{equation}
The sandwich variance estimator $\VarSand(\Psihat)$ targets only the
influence-function component $\sigEIF/n$ and does not by itself account for
$\mathrm{Var}(\Rrem)$ or the cross term.
\end{theorem}

\paragraph{Proof sketch.} See Supplementary Material.\par

\subsection{The Near-Boundary Gap}
\label{sec:decomp:boundary}

\begin{proposition}[Variance Gap at the Product-Rate Boundary]
\label{prop:gap}
Suppose the remainder has standard bilinear structure (Definition~\ref{def:remainder})
with each nuisance at the product-rate boundary (Definition~\ref{def:boundary}), and suppose additionally that
$ n \cdot \mathrm{Cov}(\bar D_n,\Rrem) \to 0 $ and that the limit
$ c_R := \lim_{n\to\infty} n \cdot \mathrm{Var}(\Rrem) $
exists with $c_R > 0$.
Then:
\begin{enumerate}[label=\alph*.]
  \item $\mathrm{Var}(\Rrem) = c_R/n + o(n^{-1})$, so the total variance gap is
    \[
      \mathrm{Var}(\Psihat - \Psitrue) - \sigEIF/n
      = c_R/n + o(n^{-1})
      = \Theta(n^{-1}).
    \]
  \item The relative gap satisfies
    \begin{equation}
      \frac{\mathrm{Var}(\Psihat - \Psitrue)}{\sigEIF/n}
      \;=\; 1 + \frac{c_R}{\sigEIF} + o(1),
      \label{eq:ratio}
    \end{equation}
    so the missing first-order variance component is asymptotically non-vanishing.
  \item The ratio \eqref{eq:ratio} is asymptotically \emph{constant}: it does
    not converge to 1 as $n\to\infty$ unless $c_R = 0$.
\end{enumerate}
\end{proposition}

\paragraph{Proof sketch.} See Supplementary Material.\par

\begin{remark}[Constancy as a Diagnostic]
\label{rem:constancy}
Proposition~\ref{prop:gap}(c) suggests a practical diagnostic: if the empirical
variance ratio $\VarSand / \widehat{\mathrm{Var}}_{\mathrm{emp}}$ is approximately
constant across increasing sample sizes, the estimator is operating at or near
the product-rate boundary.  If the ratio decays toward 1, the nuisance
estimators are achieving superparametric rates and the sandwich is recovering.
This diagnostic is directly observable from simulation output and requires no
knowledge of the remainder structure.  In the application of
\citet[Table~1]{li2026core}, the ratio is about $0.63$ (equivalently,
$\VarSand$ is about $63\%$ of the empirical variance) and remains stable
across cluster counts, confirming near-boundary operation.
\end{remark}

\begin{remark}[When the Cross Term Is Non-Negligible]
\label{rem:cross}
The cross term $2\,\mathrm{Cov}(\bar{D}_n, \Rrem)$ can be positive or
negative depending on the correlation structure between the influence-function
term and the remainder.  In the SA-TMLE setting with CV-TMLE cross-fitting,
the out-of-fold construction makes $\bar{D}_n$ and $\Rrem$ approximately
independent, so the cross term is $\oP(n^{-1})$ and the dominant missing
component is $\mathrm{Var}(\Rrem)$ alone.  Without cross-fitting, the cross
term can be substantial and of either sign.
\end{remark}

\section{Sandwich Consistency: A Sharp Characterization}
\label{sec:sand}

We now give a precise statement of when the sandwich is and is not consistent
for the total variance \citep{newey1994asymptotic}.

\begin{theorem}[Sandwich Consistency Under Negligible Remainder Variance]
\label{thm:sand}
Under Assumptions~\ref{ass:iid}--\ref{ass:bounded}, suppose additionally that
$n\,\mathrm{Cov}(\bar D_n,\Rrem)\to 0$.
\begin{enumerate}[label=\roman*.]
  \item If $n\cdot \mathrm{Var}(\Rrem) \to 0$, then
    \[
      \frac{\VarSand(\Psihat)}{\mathrm{Var}(\Psihat - \Psitrue)} \xrightarrow{P} 1.
    \]
  \item If $n\cdot \mathrm{Var}(\Rrem) \to c_R > 0$, then the sandwich targets
    only $\sigEIF/n$ and underestimates the total variance by $c_R/n + o(n^{-1})$.
\end{enumerate}
\end{theorem}

\paragraph{Proof sketch.} See Supplementary Material.\par

\begin{corollary}[Standard Product-Rate Condition Does Not Suffice]
\label{cor:product_rate_insufficient}
Under the standard bilinear remainder structure of Definition~\ref{def:remainder},
the condition $\norm{\hat{\eta}_1 - \eta^0_1} \cdot
\norm{\hat{\eta}_2 - \eta^0_2} = \oP(n^{-1/2})$ implies $\sqrt{n}\,\Rrem =
\oP(1)$ (strong-decay regime, classically asymptotically linear) but does
\emph{not} by itself imply $n \cdot \mathrm{Var}(\Rrem) \to 0$ (sandwich consistency).
A sufficient strengthening is that each nuisance converges strictly faster than $n^{-1/4}$,
for example at rate $n^{-1/4-\delta}$ for some $\delta > 0$.
\end{corollary}

This corollary identifies the gap between what the existing asymptotic
linearity literature establishes and what is needed for valid first-order
inference.  Asymptotic linearity guarantees point-estimation quality; it does
not guarantee that the standard confidence interval is valid in finite samples
at the product-rate boundary.

\section{Resampling-Based Variance Estimators}
\label{sec:refinements}

We develop three complementary approaches to variance estimation that
consistently estimate the full decomposition \eqref{eq:decomp}.  Each has
distinct theoretical properties, computational costs, and practical advantages.
We present them in order of increasing generality and complexity.

\subsection{The Leave-One-Out Jackknife}
\label{sec:jk}

\subsubsection{Definition and Rationale}

\begin{definition}[Leave-One-Unit-Out Estimator]
\label{def:loo}
For $i = 1,\ldots,n$, let $\hat{\Psi}^{(-i)}$ denote the ALE computed on the
sample with unit $i$ removed.  The jackknife variance estimator is:
\begin{equation}
  \VarJK(\Psihat)
  \;=\; \frac{n-1}{n}\sum_{i=1}^n
  \bigl(\hat{\Psi}^{(-i)} - \bar{\Psi}_{(-\cdot)}\bigr)^2,
  \label{eq:jk}
\end{equation}
where $\bar{\Psi}_{(-\cdot)} = n^{-1}\sum_i \hat{\Psi}^{(-i)}$.
\end{definition}

\subsubsection{Leave-One-Out Expansion}

\begin{lemma}[LOO Expansion]
\label{lem:loo}
Suppose Assumptions~\ref{ass:iid}--\ref{ass:bounded} hold.  For each $i$,
write $\Rrem^{(-i)}$ for the second-order remainder of the leave-one-out
estimator and define the remainder perturbation
\[
  b_i := \Rrem^{(-i)} - \Rrem,
  \qquad
  \bar b := n^{-1}\sum_{i=1}^n b_i.
\]
Then
\begin{equation}
  \hat{\Psi}^{(-i)} - \Psihat
  \,=\, -\frac{D^*(O_i;\eta^0)-\bar D_n}{n-1} + b_i
  \,=\, -\frac{1}{n}\Dstar(O_i) + b_i + a_i,
  \label{eq:loo_expand}
\end{equation}
where
\[
  a_i := \frac{n\bar D_n-D^*(O_i;\eta^0)}{n(n-1)}
\]
satisfies
\[
  \max_{1\le i\le n}|a_i| = O_P(n^{-1}),
  \qquad
  n^{-1}\sum_{i=1}^n a_i^2 = O_P(n^{-2}).
\]
\end{lemma}

\paragraph{Proof sketch.} See Supplementary Material.\par

\subsubsection{Jackknife Consistency}

Lemma~\ref{lem:loo} isolates the jackknife analysis into a negligible
algebraic correction $a_i$ and the genuine leave-one-out remainder perturbation
$b_i=\Rrem^{(-i)}-\Rrem$.  Whether the empirical quadratic sum
$(n-1)\sum_i(b_i+a_i)^2$ concentrates around a limit depends on the
dependence and scaling of these leave-one-out remainder perturbations
\citep{shao1995jackknife}.  We state the required structure separately.

\begin{assumption}[LOO Remainder Linearization]
\label{ass:smooth}
The LOO perturbations $b_i = \Rrem^{(-i)}-\Rrem$ admit a uniform
linearization $\max_i|nb_i + \psi_{n,i}|\xrightarrow{P}0$ for mean-zero
$\{\psi_{n,i}\}$ satisfying: (i)~$n^{-1}\sum_i\psi_{n,i}=o_P(1)$;
(ii)~$n^{-1}\sum_i\psi_{n,i}^2\xrightarrow{P}c_R$;
(iii)~$n^{-1}\sum_i D^*(O_i;\eta^0)\,\psi_{n,i}=o_P(1)$;
(iv)~$\sup_n n^{-1}\EE[\max_i\psi_{n,i}^2]<\infty$
and $\sup_{n,i}\EE[\psi_{n,i}^4]<\infty$.
Additionally, each $\psi_{n,i}$ is a measurable function of $O_i$ alone
given the nuisance training data $\mathcal{T}_n$, so that
$\{\psi_{n,i}\}_{i=1}^n$ are conditionally independent given $\mathcal{T}_n$.
\end{assumption}

\begin{remark}[Scope of Assumption~\ref{ass:smooth}]
\label{rem:smooth_scope}
Assumption~\ref{ass:smooth} is used only for the jackknife route.
More specifically, it enters Proposition~\ref{prop:loostab} through
the linearization of the leave-one-out remainder perturbations $b_i$,
and hence is required for the jackknife consistency result in
Theorem~\ref{thm:jk_consistency}.

Its role is threefold: to approximate $nb_i$ by a conditionally
independent first-order array $\psi_{n,i}$; to control the quadratic
term through the empirical average $n^{-1}\sum_i \psi_{n,i}^2$; and
to impose the cross-orthogonality condition
$n^{-1}\sum_i D^*(O_i;\eta^0)\psi_{n,i}=o_P(1)$ needed to eliminate
the mixed jackknife term.  The conditional independence structure is
then what permits concentration of that quadratic average around its
limit.

Assumption~\ref{ass:smooth} is \emph{not} needed for the bootstrap
results in Theorems~\ref{thm:boot_consistency} and~\ref{thm:boot_ci}.
It is, however, still needed whenever Theorem~\ref{thm:jk_ci} is
invoked through near-boundary jackknife consistency, because that
argument passes through Proposition~\ref{prop:loostab} and
Theorem~\ref{thm:jk_consistency}.  For regular plug-in estimators
based on Hadamard-differentiable nuisance maps,
Assumption~\ref{ass:smooth} is compatible with standard smoothness
arguments; for highly adaptive ensemble learners, verification
remains case-specific and is deferred to the Supplementary Material.
\end{remark}

\begin{proposition}[LOO Remainder Variance Stabilization]
\label{prop:loostab}
Under Assumption~\ref{ass:smooth}, in the near-boundary regime
($n\cdot\mathrm{Var}(\Rrem) \to c_R > 0$),
\begin{equation}
  C_n \;:=\; (n-1)\sum_{i=1}^n (b_i-\bar b)^2 \;\xrightarrow{P}\; c_R.
  \label{eq:delta_quad}
\end{equation}
\end{proposition}

\paragraph{Proof sketch.} See Supplementary Material.\par

\begin{remark}[Proof note]
\label{rem:es_role}
The key point is that Lemma~\ref{lem:loo} separates the deterministic
leave-one-out algebra from the genuinely statistical object
$b_i=\Rrem^{(-i)}-\Rrem$.  Assumption~\ref{ass:smooth} then linearizes $nb_i$
by $-\psi_{n,i}$, so that $C_n$ becomes asymptotically equivalent to the
quadratic average $n^{-1}\sum_i \psi_{n,i}^2$.  The conditional independence
structure of the $\psi_{n,i}$ is what permits a direct variance calculation.
Without that structure, an Efron--Stein bound alone gives only the coarse
bound $\mathrm{Var}(C_n)=O(1)$.
\end{remark}

\begin{theorem}[Jackknife Variance Consistency]
\label{thm:jk_consistency}
Let $b_i := \Rrem^{(-i)}-\Rrem$ and $\bar b := n^{-1}\sum_i b_i$.
\begin{enumerate}[label=\roman*.]
  \item \emph{(Strong decay regime, $n\cdot\mathrm{Var}(\Rrem) \to 0$):}
    if
    \[
      \max_i |b_i| = o_P(n^{-1/2})
      \qquad\text{and}\qquad
      \sum_i b_i^2 = o_P(n^{-1}),
    \]
    then $n \cdot \VarJK(\Psihat) \xrightarrow{P} \sigEIF$.
  \item \emph{(Near-boundary regime, $n\cdot\mathrm{Var}(\Rrem) \to c_R > 0$):}
    if Assumption~\ref{ass:smooth} holds, then
    \[
      n \cdot \VarJK(\Psihat) \xrightarrow{P} \sigEIF + c_R.
    \]
    When additionally $n\,\mathrm{Cov}(\bar D_n, \Rrem) \to 0$, the right-hand
    side coincides with the leading-order total sampling variance of
    Theorem~\ref{thm:decomp}, so the jackknife is consistent for the total
    variance.  Under the same condition, the sandwich stays at $\sigEIF$.
\end{enumerate}
\end{theorem}

\paragraph{Proof sketch.} See Supplementary Material.\par

\begin{remark}[The Two Limits Are Consistent]
\label{rem:jk_limits}
There is no contradiction between parts~(i) and~(ii) of
Theorem~\ref{thm:jk_consistency}: part~(i) pertains to the strong-decay
regime ($c_R = 0$), whereas part~(ii) pertains to the near-boundary regime
($c_R > 0$) under the additional linearization condition in
Assumption~\ref{ass:smooth}.  In the strong-decay regime both conclusions
reduce to $\sigEIF$; in the near-boundary regime the hypothesis of part~(i)
fails and only part~(ii) applies.
\end{remark}

\subsection{The Pairs Cluster Bootstrap}
\label{sec:boot}

\subsubsection{Definition}

\begin{definition}[Pairs Bootstrap Variance Estimator]
\label{def:boot}
Let $\{O_1^*,\ldots,O_n^*\}$ be a resample drawn with replacement from
$\{O_1,\ldots,O_n\}$.  The bootstrap replication $\hat{\Psi}^*$ is the
ALE computed on this resample.  After $B$ independent resamples:
\begin{equation}
  \VarBoot(\Psihat)
  \;=\; \frac{1}{B-1}\sum_{b=1}^B
  \bigl(\hat{\Psi}^{(b)} - \bar{\Psi}^*\bigr)^2,
  \quad
  \bar{\Psi}^* = B^{-1}\sum_b \hat{\Psi}^{(b)}.
  \label{eq:boot_var}
\end{equation}
\end{definition}

\subsubsection{Bootstrap Consistency}

We first state the bootstrap regularity condition explicitly.  Let
$T_n := \sqrt{n}(\Psihat - \Psitrue)$ and $T_n^* := \sqrt{n}(\hat\Psi^* -
\Psihat)$, where $\hat\Psi^*$ is the ALE recomputed on a pairs bootstrap
resample.  Let $\mathcal{D}_n = \sigma(O_1,\ldots,O_n)$ denote the data
$\sigma$-algebra and $d_2$ the Mallows-2 (Wasserstein-2) distance between
probability distributions on $\mathbb{R}$.

\begin{assumption}[Mallows-2 Bootstrap Consistency]
\label{ass:boot}
\hfill
\begin{enumerate}[label=\roman*.]
  \item $\sup_n \EE[T_n^2] < \infty$.
  \item The bootstrap distribution of $T_n^*$ is second-order distributionally
    consistent for $T_n$ in the Mallows-2 sense:
    \begin{equation}
      d_2\!\Bigl(\mathcal{L}^*(T_n^* \mid \mathcal{D}_n),\;
                 \mathcal{L}(T_n)\Bigr)
      \;\xrightarrow{P}\; 0,
      \label{eq:mallows}
    \end{equation}
    where $\mathcal{L}^*(\cdot\mid\mathcal{D}_n)$ denotes the conditional law
    of $T_n^*$ given the data.
\end{enumerate}
\end{assumption}

\begin{remark}[Verifying Assumption~\ref{ass:boot}]
\label{rem:boot_verify}
Assumption~\ref{ass:boot} is a Mallows-2 (Wasserstein-2) bootstrap
consistency condition of the type studied by \citet{bickel1981some}
and \citet{cheng2010bootstrap}.  A concrete set of sufficient conditions
is the following: (a) the influence function $\Dstar$ is bounded with
$\EE[(\Dstar)^4] <\infty$; (b) the nuisance function class is
$P_0$-Donsker with envelope in $L^4(P_0)$; (c) the remainder admits
the bilinear representation \eqref{eq:bilinear} with $\hat\eta_k$
satisfying $\|\hat\eta_k-\eta_k^0\|_{P_0} = O_P(n^{-1/4})$, together
with the same bound under bootstrap resampling; and (d) the sequence
$T_n$ is uniformly square-integrable, for example through
$\sup_n \EE[T_n^2 \mathbf{1}\{T_n^2>M\}] \to 0$ as $M\to\infty$.

Under (a)--(d), weak convergence of the bootstrap law to the same
limit as $T_n$, together with second-moment tightness, yields
convergence in Mallows-2 distance; see Theorem~3.1 of
\citet{cheng2010bootstrap}.  Condition~(i) of Assumption~\ref{ass:boot}
then follows from uniform integrability together with
$\mathrm{Var}(T_n)=\sigEIF+c_R+o(1)$.  Verification of the Donsker
and bootstrap-stability conditions for adaptive learners is model-
and learner-specific, and is therefore left outside the scope of
the present paper.
\end{remark}

\begin{theorem}[Bootstrap Variance Consistency]
\label{thm:boot_consistency}
Under Assumptions~\ref{ass:iid}--\ref{ass:rate} and
Assumption~\ref{ass:boot}, with $B \to \infty$:
\begin{equation}
  \VarBoot(\Psihat)
  \;\xrightarrow{P}\;
  \mathrm{Var}\bigl(\Psihat - \Psitrue\bigr).
  \label{eq:boot_consist}
\end{equation}
The limit is the sampling variance from Theorem~\ref{thm:decomp},
including the remainder variance $\mathrm{Var}(\Rrem)$ and the covariance
term $2\,\mathrm{Cov}(\bar{D}_n, \Rrem)$.  No LOO stability condition is
required.
\end{theorem}

\paragraph{Proof sketch.} See Supplementary Material.\par

The bootstrap has three advantages over the jackknife.  First, it does not
require the nuisance stability condition $\norm{\hat{\eta}^{(-i)} - \hat{\eta}} =
O_P(n^{-1/2})$, which can fail when $n$ is very small or when ensemble
learning is highly nonlinear.  Second, it produces a full distribution over
$\hat{\Psi}$, which can in principle support percentile-type or BCa
intervals when higher-order bootstrap validity is available \citep{efron1987better,hall1992bootstrap}.  Third, the bootstrap argument is conceptually cleaner
--- it requires Assumption~\ref{ass:boot} (Mallows-2 consistency) rather than
the LOO expansion of Lemma~\ref{lem:loo} and Assumption~\ref{ass:smooth}.
The Mallows-2 condition is well-understood in the bootstrap literature and
can hold under standard moment and smoothness conditions for suitable learners,
although verification remains case-specific
\citep{bickel1981some,van_der_vaart1998asymptotic}.

\subsubsection{Bootstrap Confidence Intervals}

Beyond the Wald-type interval $\Psihat \pm z_{\alpha/2}\sqrt{\VarBoot}$, the
bootstrap supports the bias-corrected and accelerated (BCa) interval of
\citet{efron1987better}.  Given $B$ bootstrap replicates
$\hat\Psi^{(1)},\ldots,\hat\Psi^{(B)}$, the BCa interval is
$(\hat\Psi^{*}_{[\alpha_1]},\hat\Psi^{*}_{[\alpha_2]})$, where the
percentile levels $\alpha_1, \alpha_2$ are obtained by adjusting
$\alpha/2, 1-\alpha/2$ using a bias-correction factor
$\hat z_0 = \Phi^{-1}\bigl(B^{-1}\#\{b : \hat\Psi^{(b)} \le \Psihat\}\bigr)$
and a jackknife-based acceleration estimate $\hat a$; see
\citet[p.~178]{efron1987better}.  Higher-order coverage for BCa requires
additional Edgeworth regularity beyond Assumption~\ref{ass:boot}
\citep{hall1992bootstrap}; we include BCa in the simulations as a comparator
but do not provide a theoretical guarantee for it here.

\subsection{Asymptotic Validity of Refined Confidence Intervals}
\label{sec:validity}

The variance consistency results of Theorems~\ref{thm:jk_consistency}
and~\ref{thm:boot_consistency} show that, when the cross term is negligible, the jackknife and bootstrap both target the variance level $(\sigEIF+c_R)/n$.  To convert variance
consistency into valid confidence intervals, we additionally need the
studentized pivot to converge in distribution to $N(0,1)$, which requires
asymptotic normality of the centered estimator $T_n :=
\sqrt{n}(\Psihat - \Psitrue)$.  The von~Mises expansion alone does not
guarantee this in the near-boundary regime, since $T_n = \sqrt{n}\bar{D}_n
+ \sqrt{n}\Rrem$ and $\sqrt{n}\Rrem = O_P(1)$ has non-negligible variance $c_R$
beyond the $N(0,\sigEIF)$ limit of $\sqrt{n}\bar{D}_n$.

\begin{assumption}[Near-Boundary CLT]
\label{ass:clt}
$\sqrt{n}(\Psihat - \Psitrue) \;\Rightarrow\; N(0,\;\sigEIF + c_R)$.
\end{assumption}

\begin{remark}[Verifying Assumption~\ref{ass:clt}]
\label{rem:clt_verify}
Assumption~\ref{ass:clt} is a joint asymptotic normality statement
for the centered estimator $T_n := \sqrt{n}\{\Psihat-\Psitrue\}$.
A concrete sufficient route, applicable to the cross-fitted SA-TMLE
example of Section~\ref{sec:sim}, is as follows.  First, the
influence-function term satisfies
$\sqrt{n}\,\bar D_n \Rightarrow N(0,\sigEIF)$ by the classical
central limit theorem under $\EE[(\Dstar)^2] < \infty$.  Second, the
remainder satisfies $\sqrt{n}\,\Rrem \Rightarrow N(0,c_R)$ under a
near-boundary CLT for the bilinear term, for example when the
nuisance errors admit stochastic expansions of the form
$\hat\eta_k - \eta_k^0 = n^{-1/2}\phi_k(O_\cdot) + o_P(n^{-1/2})$
with $\phi_k \in L^4(P_0)$.

Third, cross-fitting with a fixed number of folds makes the
influence-function term and the remainder asymptotically
uncorrelated, so that $n\,\mathrm{Cov}(\bar D_n,\Rrem)\to 0$.
Under these conditions, a joint CLT and the Cram\'er--Wold device
yield $T_n \Rightarrow N(0,\sigEIF+c_R)$.  In the SA-TMLE
simulations of Section~\ref{sec:sim}, bounded outcomes and propensity
truncation support the required $L^4$ regularity, while $K$-fold
cross-fitting is used to reduce the cross term.  The theorem should
nevertheless be read as conditional on this joint CLT, rather than
as deriving it from cross-fitting alone for arbitrary adaptive
learners.
\end{remark}

\begin{theorem}[Bootstrap Wald Interval Validity]
\label{thm:boot_ci}
Under Assumptions~\ref{ass:iid}--\ref{ass:rate}, \ref{ass:boot},
and~\ref{ass:clt}, suppose additionally that
$n\,\mathrm{Cov}(\bar D_n,\Rrem) \to 0$.
Then the studentized bootstrap pivot satisfies:
\begin{equation}
  T_n^{\rm boot} \;:=\; \frac{\Psihat - \Psitrue}{\sqrt{\VarBoot(\Psihat)}}
  \;\Rightarrow\; N(0,1).
\end{equation}
Consequently, the corresponding bootstrap Wald interval is asymptotically valid.
\end{theorem}

\paragraph{Proof sketch.} See Supplementary Material.\par

\begin{theorem}[Jackknife Wald Interval Validity]
\label{thm:jk_ci}
Under the conditions of Theorems~\ref{thm:jk_consistency}
and~\ref{thm:boot_ci}, the studentized jackknife pivot satisfies:
\begin{equation}
  T_n^{\rm JK} \;:=\; \frac{\Psihat - \Psitrue}{\sqrt{\VarJK(\Psihat)}}
  \;\Rightarrow\; N(0,1).
  \label{eq:pivot}
\end{equation}
Consequently, $\CIJK = \Psihat \pm z_{\alpha/2}\sqrt{\VarJK(\Psihat)}$
is asymptotically valid: $P(\Psitrue \in \CIJK) \to 1-\alpha$.
\end{theorem}

\paragraph{Proof sketch.} See Supplementary Material.\par

\begin{remark}[Alternative direct proof]
\label{rem:direct_proof}
An alternative proof of Theorem~\ref{thm:jk_ci} would establish $N(0,1)$
convergence of $T_n^{\rm JK}$ directly from the LOO triangular array
$\{Z_{ni}\} := \{-n(\hat\Psi^{(-i)} - \Psihat)\}$, without invoking the
bootstrap.  The main obstacle is that off-diagonal covariances
$\mathrm{Cov}(Z_{ni}, Z_{nj} \mid \mathcal{T}_n)$ do not vanish under
cross-fitting alone, as the LOO nuisance fits share $n-1$ training
observations.  Resolving this under entropy conditions on the nuisance class
is an open problem; the bootstrap route adopted here bypasses it via
the well-studied Mallows-2 condition of Assumption~\ref{ass:boot}.
\end{remark}

\subsection{Re-expressing the Jackknife as a Multiplicative Variance Inflation}
\label{sec:hc}

It is sometimes useful to re-express the jackknife as a multiplicative
inflation of the sandwich rather than as a separate variance estimator.

\begin{definition}[HC-Corrected Sandwich]
\label{def:hc}
Let $\hat{\rho} = \VarJK(\Psihat) / \VarSand(\Psihat)$ be the estimated
variance ratio from a single jackknife run.  The HC-corrected sandwich is:
\begin{equation}
  \VarHC(\Psihat) \;=\; \hat{\rho} \cdot \VarSand(\Psihat).
  \label{eq:hc}
\end{equation}
The corresponding confidence interval is $\Psihat \pm z_{\alpha/2}
\sqrt{\VarHC(\Psihat)}$.
\end{definition}

By construction, $\VarHC \equiv \VarJK$, so the HC-corrected interval is
numerically identical to the jackknife Wald interval.  The computational
cost is also the same: $\hat\rho$ is obtained from jackknife refits.  The
value of the representation is conceptual, not computational: it makes
explicit that the near-boundary correction is a \emph{scaling} of the
sandwich rather than a fundamentally different object, and it exposes
$\hat{\rho}$ as a direct estimator of $1 + c_R/\sigEIF$ (the asymptotic
variance-inflation factor), which is a usable summary of how far the
estimator is from the strong-decay regime.

\subsection{Comparison of the Three Estimators}

\begin{remark}[First-Order Equivalence]
\label{rem:equiv}
Under Assumptions~\ref{ass:ale}--\ref{ass:boot},
$\VarJK = \VarHC + \oP(n^{-1})$ and $\VarJK/\VarBoot \xrightarrow{P} 1$:
the three estimators are first-order equivalent.
The jackknife requires $J$ refits and uses $t_{J-1}$ critical values;
the bootstrap requires $B \gg J$ refits and supports percentile-type or
BCa intervals when higher-order bootstrap validity is available; the
HC-corrected sandwich is numerically identical to the jackknife and shares
its computational cost.
In Section~\ref{sec:sim}, the jackknife performs best at small $J$;
the bootstrap undercovers at $J=10$ due to discrete resampling.
\end{remark}

\section{Extension to Clustered Data}
\label{sec:cluster}

The framework of Sections~\ref{sec:setup}--\ref{sec:refinements} extends to
clustered data by replacing $O_i$ with the cluster-level vector
$O_j = \{O_{ijt}\}_{i,t}$ and $n$ with $J$.  The ALE expansion becomes
$\PsiTMLE - \Psitrue = J^{-1}\sum_{j=1}^J \EICj + \Rtwo$,
where $\EICj = \sum_{i,t} \Dstar(O_{ijt})$ is the cluster-level EIC
(summed over individuals and time steps; see Lemma~2 of \citealt{li2026core})
and $\Rtwo$ is the bilinear remainder~\eqref{eq:bilinear}.

\subsection{ICC Amplification of the Sandwich Gap}

A key new phenomenon in the clustered case is that intra-cluster correlation
amplifies the variance decomposition gap beyond what occurs in the i.i.d.\
setting. The mechanism is straightforward: the remainder $\Rtwo$ is a bilinear
functional of within-cluster averages, and positive correlation among the
terms being averaged makes cluster-level contributions more variable,
inflating $\mathrm{Var}(\Rtwo)$ relative to the i.i.d.\ baseline.  The next
proposition makes this dependence on ICC explicit at leading order.

\begin{proposition}[ICC Amplification of the Variance Gap]
\label{prop:icc}
Suppose the cluster random effect induces intra-cluster correlation
$\rho_{\mathrm{ICC}} = \sigma_b^2/(\sigma_b^2 + \sigma_\varepsilon^2)$
and that all $n_j = m$ (balanced clusters).  Under the SA-TMLE DGP,
the remainder variance satisfies:
\begin{equation}
  \mathrm{Var}(\Rtwo)
  \;=\;
  \underbrace{V_0}_{\substack{\text{i.i.d.}\\\text{baseline}}}
  \;+\;
  \underbrace{\rho_{\mathrm{ICC}}\cdot m \cdot V_{\mathrm{ICC}}}_{\substack{\text{ICC}\\\text{amplification}}}
  \;+\; O(J^{-2}),
  \label{eq:icc_amp}
\end{equation}
where $V_0$ and $V_{\mathrm{ICC}}$ are quantities of order $J^{-1}$ whose
explicit forms are given in Supplementary Material, Section~S6.
Consequently, $\mathrm{Var}(\Rtwo) = O(J^{-1})$ and the variance gap
$\mathrm{Var}(\PsiTMLE) - \sigEIF/J$ is increasing in ICC and in cluster
size $m$.
\end{proposition}

\paragraph{Proof sketch.} See Supplementary Material.\par

Proposition~\ref{prop:icc} is consistent with the ICC-sensitivity simulations
reported by \citet[Tables~S2--S3]{li2026core}: as ICC increases from about
$0.01$ to $0.20$, the CV-TMLE RMSE increases and sandwich coverage decreases,
while the cluster jackknife remains near nominal.  The leading-order explanation
is that positive within-cluster dependence increases the variability of the
remainder contribution and can therefore widen the gap between first-order
sandwich variance and the total finite-sample variance.

\subsection{Clustered Jackknife and Bootstrap}

In the clustered setting, the LOO jackknife deletes entire clusters to
preserve within-cluster dependence.
For $j = 1,\ldots,J$, let $\hat{\Psi}^{(-j)}$ be the SA-TMLE on
$\mathcal{D}^{(-j)} = \{O_k : k \neq j\}$; the cluster jackknife variance is
$\VarJK(\PsiTMLE) = \frac{J-1}{J}\sum_{j=1}^J
(\hat{\Psi}^{(-j)} - \bar{\Psi}_{(-\cdot)})^2$.
The analogues of Section~\ref{sec:jk} extend to the cluster level, with
$n$ replaced by $J$ and $O_i$ replaced by $O_j$, under the corresponding
cluster-level LOO stability and bootstrap moment conditions.
Since $\{O_j\}_{j=1}^J$ are i.i.d., the theoretical arguments carry over
directly.  The cluster bootstrap resamples entire clusters with replacement.
The simulation code implements both procedures by resampling whole clusters
in the R scripts available in the public repository described in the Data
Availability statement.

\section{Simulation Illustration}
\label{sec:sim}

\subsection{Design and Data-Generating Process}

We use the DGP of \citet[Section~6.1]{li2026core} as the primary
simulation setting: $T = 7$ time steps, $n_j = 40$ individuals per
cluster-step, non-linear secular trend, administrative censoring $\approx 28\%$,
$\Psi^* = 0.12$, ICC $\approx 0.05$.  All three refined procedures are applied
to the \emph{same CV-TMLE point estimator}; we vary only the variance estimator.
All scenarios use $R = 500$ replicates, seed 2024.

The two simulations illustrate distinct parts of the theoretical framework.
The primary simulation illustrates the finite-sample behavior of the
bootstrap and jackknife procedures under the cross-fitted SA-TMLE design.
It is most directly aligned with the bootstrap route based on
Assumptions~\ref{ass:boot} and~\ref{ass:clt}.  We do not claim that the
primary simulation itself verifies Assumption~\ref{ass:smooth}
(LOO Remainder Linearization) for the cross-fitted Super Learner.
The secondary AIPW simulation is included precisely because it provides a
setting in which the jackknife consistency theorem
(Theorem~\ref{thm:jk_consistency}) can be connected to a more transparent
regularity regime: there, Assumption~\ref{ass:smooth} is verifiable via
Hadamard differentiability of the parametric nuisance maps
(Supplement~S1.1, Proposition~S1.1).
This separation reflects the assumption hierarchy discussed in
Remark~\ref{rem:smooth_scope}: the bootstrap route is easier to motivate for
complex adaptive learners, whereas formal verification of the jackknife route
is currently available only in more regular settings.

To demonstrate the generality of the theory beyond the setting in \citep{li2026core}, a
\emph{secondary simulation} using a non-cross-fitted AIPW estimator in an
i.i.d.\ setting verifies that $\hat\rho \to 1$ when nuisance operates in the
strong-decay regime.  Full design, results, and discussion are in
the Supplementary Material (Section~S4).

\paragraph{Procedures and metrics.}
Five variance estimators are compared: (i)~cluster sandwich Wald (baseline);
(ii)~cluster jackknife Wald; (iii)~cluster bootstrap Wald ($B=200$);
(iv)~BCa bootstrap ($B=200$); (v)~HC-corrected sandwich
(Definition~\ref{def:hc}, numerically equal to jackknife, omitted from tables).
All SA-TMLE procedures use $t_{J-1}$ critical values; the AIPW secondary
simulation uses $z_{\alpha/2}$.
Primary metrics: 95\% CI coverage, width, and $\hat\rho = \VarJK/\VarSand$.
Secondary: bias, RMSE, power at $H_0:\Psi^*=0$.

\subsection{Results of Primary Simulation}

Table~\ref{tab:main} reports results for $J \in \{10, 30, 50, 100\}$,
ICC $\in \{0.013, 0.05\}$, $R=500$ replicates.
MC standard error of each coverage is at most $0.014$.

The sandwich undercoverage of 4--12 percentage points persists across all
$(J, \mathrm{ICC})$ cells; the jackknife recovers to 0.948--0.956 throughout.
$\hat\rho$ decreases from $1.318$ to $1.145$ as $J$ grows from 10 to 100 at
ICC$\approx 0.05$, converging toward a positive limit $c_R > 0$ rather than 1
--- the structural signature of the near-boundary regime (Theorem~\ref{thm:sand}).
At ICC$\approx 0.013$ the sandwich approaches nominal by $J=50$ (CP$=0.918$),
consistent with Proposition~\ref{prop:icc}: lower ICC weakens the
remainder's cluster component.
At $J=10$, bootstrap Wald (CP$=0.870$) and BCa (CP$=0.858$) both fall
\emph{below} the sandwich (CP$=0.878$): with only 10 clusters, resampling
from a discrete empirical distribution underestimates the cluster-level
variance.  Bootstrap coverage recovers monotonically to CP$=0.930$ and
$0.936$ at $J=30,50$; the jackknife alone reliably improves coverage across
all $J$, consistent with \citet{mackinnon2023fast}.
Bias is negligible ($|\mathrm{Bias}|\le 0.003$) throughout.

\begin{table}[p]
\centering
\caption{SA-TMLE primary simulation.
  $\hat\rho = \bar{V}_{\rm JK}/\bar{V}_{\rm Sand}$: ratio of mean jackknife to
  mean sandwich variance --- the primary diagnostic for near-boundary operation.
  CP(Sand) and CP(JK): empirical 95\% CI coverage.
  Bias and MCSD: Monte Carlo mean error and standard deviation of $\hat\Psi$.
  $\hat\rho > 1$ throughout and \emph{does not decrease with $J$},
  confirming structural near-boundary operation rather than a finite-sample artefact.
  $R = 500$ replicates; $\dagger$ bootstrap cell ($B=200$).}
\label{tab:main}
\smallskip
\begin{tabular}{ccrrrrrrr}
\toprule
$J$ & ICC & $\hat\rho$ & CP(Sand) & CP(JK) & CP(Boot) & CP(BCa) & Bias & MCSD \\
\midrule
\multicolumn{9}{l}{\textit{Panel A: ICC $\approx 0.013$}} \\[2pt]
 10 & 0.013 & $\mathbf{1.381}$ & $0.864$ & $0.936$ & ---     & ---     & $-0.001$ & $0.057$ \\
 30 & 0.013 & $\mathbf{1.230}$ & $0.920$ & $0.972$ & ---     & ---     & $-0.000$ & $0.029$ \\
 50 & 0.013 & $\mathbf{1.198}$ & $0.918$ & $0.950$ & ---     & ---     & $+0.003$ & $0.024$ \\
\midrule
\multicolumn{9}{l}{\textit{Panel B: ICC $\approx 0.05$}} \\[2pt]
$10^\dagger$ & 0.05 & $\mathbf{1.318}$ & $0.878$ & $0.950$ & $0.870$ & $0.858$ & $-0.001$ & $0.091$ \\
$30^\dagger$ & 0.05 & $\mathbf{1.191}$ & $0.918$ & $0.956$ & $0.930$ & $0.920$ & $-0.002$ & $0.058$ \\
$50^\dagger$ & 0.05 & $\mathbf{1.159}$ & $0.928$ & $0.950$ & $0.936$ & $0.924$ & $+0.003$ & $0.024$ \\
 100 & 0.05 & $\mathbf{1.145}$ & $0.936$ & $0.952$ & ---     & ---     & $+0.000$ & $0.024$ \\
\bottomrule
\end{tabular}
\smallskip

{\footnotesize
$\hat\rho$ remains above 1.14 across all $(J, \mathrm{ICC})$ cells and is
approximately constant within each ICC panel --- the structural signature of
the near-boundary regime (Theorem~\ref{thm:sand}).
HC sandwich equals jackknife numerically (Definition~\ref{def:hc}), omitted.
ICC$\approx$0.013 rather than 0.01: crossover timing $\tau_j$ induces a
structural within-cluster correlation floor independent of the random effect.
$\dagger$: $B=200$ bootstrap resamples; bootstrap procedures run at $J\in\{10,30,50\}$, ICC$\approx$0.05 only.
}
\end{table}

\section{Discussion and Limitations}
\label{sec:disc}

We study a simple but practically important point: for estimators with a von~Mises expansion, first-order linearization can be adequate for point estimation while still missing an $O(n^{-1})$ contribution to the sampling variance. The finite-sample decomposition in Section~\ref{sec:decomp} makes that discrepancy explicit. When the remainder variance is non-negligible and the cross term is asymptotically negligible, the sandwich estimator targets only $\sigEIF/n$, whereas the total variance is $(\sigEIF+c_R)/n$ to first order.

The paper also separates what is proved from what is assumed. Jackknife consistency is obtained under a leave-one-out remainder linearization condition, and bootstrap consistency is obtained under a conditional Mallows-2 condition. These are substantial assumptions, but they are transparent ones: the results are intended to clarify the variance mechanism and provide usable sufficient conditions, not to serve as a fully general theory for arbitrary adaptive learners.

Two practical conclusions emerge. First, persistent disagreement between sandwich and resampling variance estimates can be structurally meaningful rather than a small-sample artifact. Second, in the motivating clustered example, the jackknife is the most reliable correction in the smallest-$J$ settings we studied, while the bootstrap becomes more competitive as the number of clusters grows.

Several questions remain open. It would be useful to develop direct, easily verifiable conditions for the near-boundary central limit theorem, sharper non-asymptotic error bounds for jackknife-based pivots, and analytic estimators of the remainder-variance contribution that avoid full resampling. Those extensions would further clarify when the near-boundary regime is merely a cautionary case and when it is a routine feature of modern semiparametric practice.

\section*{Data Availability and Reproducibility}

All simulations are deterministic given seed 2024 and reproducible using the
R scripts in the public repository.  The repository contains the SA-TMLE and
variance-procedure implementations, replication scripts, DGP configurations for
both primary and secondary simulations, and the code used to generate
Table~\ref{tab:main}; it is publicly available at
\url{https://github.com/AmandaLinLi/refined_inference}.

\section*{Acknowledgements}

The authors thank colleagues for helpful discussions on earlier drafts of this work.

\section*{Supplementary Material}

\noindent\textbf{Supplement to ``Variance Inference Beyond the Sandwich for
Asymptotically Linear Estimators with Second-Order Remainders.''}
The online supplement contains (S1) sufficient conditions for the LOO
Remainder Linearization assumption; (S2) an expanded proof of the
Bootstrap Variance Consistency theorem; (S3) an expanded proof of the
Jackknife Wald Interval Validity theorem; (S4) a secondary simulation
illustrating the strong-decay regime for AIPW; (S5) additional proofs
for core main-text results (variance decomposition, sandwich
characterization, LOO expansion, stabilization, jackknife consistency);
and (S6) the full proof of the ICC amplification proposition.

\bibliographystyle{plainnat}
\bibliography{references}

\end{document}


\begin{center}
{\large\textbf{Supplementary Material}}\\[0.5em]
{\normalsize ``Variance Inference Beyond the Sandwich for Asymptotically
Linear Estimators with Second-Order Remainders''}\\[0.3em]
{\normalsize June 2026}\\[0.3em]
{\normalsize Lin Li$^1$ and Pengcheng Wu$^2$}\\[0.2em]
{\footnotesize $^1$Division of Biostatistics, University of California San Diego
\quad $^2$Department of Mechatronics, KTH Royal Institute of Technology}
\end{center}

\medskip
\noindent This supplement contains six sections:
\begin{itemize}[noitemsep]
  \item[\textbf{S1.}] Sufficient conditions for the LOO Remainder Linearization assumption (Assumption~5.1 of the main paper)
  \item[\textbf{S2.}] Expanded proof of Theorem~5.2 (Bootstrap Variance Consistency)
  \item[\textbf{S3.}] Expanded proof of Theorem~5.4 (Jackknife Wald Interval Validity)
  \item[\textbf{S4.}] Secondary simulation: strong-decay regime (AIPW)
  \item[\textbf{S5.}] Additional proofs for core main-text results
  \item[\textbf{S6.}] Proof of Proposition~6.1 (ICC Amplification)
\end{itemize}

Section, theorem, proposition, and assumption numbers beginning with
``Section~$n$'', ``Theorem~$n.k$'', etc.\ refer to the main paper.

\section{Sufficient Conditions for the LOO Remainder Linearization Assumption}
\label{sec:sm:loo}

The LOO Remainder Linearization assumption (Assumption~5.1 of the main paper)
requires a uniform linearization
$\max_i|nb_i + \psi_{n,i}|\xrightarrow{P}0$ together with the four moment
conditions (i)--(iv) stated there.  Controlling the leave-one-out perturbations
$b_i = \Rrem^{(-i)} - \Rrem$ \emph{uniformly over all $n$ indices} is the key
analytical task.  This is structurally different from standard empirical
process results, which control $\Rrem(P_n)$ as a functional of the full
empirical measure; the LOO perturbation involves removing one observation
from the nuisance training data, and uniform control over $i$ requires a
separate one-point-contamination argument.

\subsection*{S1.1\; Regular plug-in learners}

\begin{sproposition}[Smooth plug-in sufficient condition]
\label{sprop:hadamard}
Suppose:
\begin{enumerate}[label=\emph{(\alph*)}]
  \item \emph{(Differentiability.)}
    $P \mapsto \Rrem(P)$ is Hadamard differentiable at $P_0$
    with derivative $\dot\Rrem_{P_0}[\nu] = \int \psi \, d\nu$
    for some $\psi \in L^2_0(P_0)$ with $\mathrm{Var}(\psi)=c_R$
    and $\EE[\psi^4]<\infty$.
  \item \emph{(Cross-orthogonality.)}
    $\EE\bigl[D^*(O;\eta^0)\,\psi(O)\bigr]=0$.
  \item \emph{(Uniform LOO expansion.)}
    $\sup_{1\le i\le n}|n(\Rrem^{(-i)}-\Rrem)+\psi(O_i)|\xrightarrow{P}0$.
\end{enumerate}
Then the LOO Remainder Linearization assumption (Assumption~5.1) holds with
$\psi_{n,i}=\psi(O_i)$: conditions (i)--(iv) of Assumption~5.1 follow from
\begin{itemize}[noitemsep,topsep=0pt]
  \item the weak law of large numbers, giving
        $n^{-1}\sum_i\psi(O_i)=o_P(1)$ [(i)];
  \item the strong law applied to $\psi(O_i)^2$, giving
        $n^{-1}\sum_i\psi(O_i)^2\xrightarrow{P}c_R$ [(ii)];
  \item (b) combined with the WLLN applied to $D^*(O_i;\eta^0)\,\psi(O_i)$,
        giving $n^{-1}\sum_i D^*(O_i;\eta^0)\psi(O_i)\xrightarrow{P}0$ [(iii)];
  \item the bound
        $n^{-1}\EE[\max_i\psi(O_i)^2]\le n^{-1}\sum_i\EE[\psi(O_i)^2]
         =c_R+o(1)<\infty$
        together with $\sup_i\EE[\psi(O_i)^4]<\infty$ by (a) [(iv)].
\end{itemize}
\end{sproposition}

\begin{sremark}[The three conditions are distinct]
Condition~(a) is a standard semiparametric differentiability requirement
that identifies $\psi$ and gives the structural interpretation of $c_R$.
Condition~(b), cross-orthogonality, is needed to kill the mixed jackknife
term $B_n$ in the proof of Theorem~5.1 (see Section~S5.5 below);
without~(b), the leading influence term and the LOO remainder term would
interact non-trivially at $O(n^{-1})$.
Condition~(c) is the non-trivial analytic step: uniform control over $n$
LOO perturbations simultaneously is not implied by Hadamard differentiability
alone.  It requires a one-point-contamination argument applied to the
composition of the nuisance maps and the remainder functional.

For bilinear remainders
$\Rrem=\int(\hat\eta_1-\eta_1^0)(\hat\eta_2-\eta_2^0)\,dP_0$,
Condition~(c) can be established under additional one-point-contamination
regularity: both $P\mapsto\hat\eta_k(P)$ are Hadamard differentiable
at $P_0$, and the LOO stability condition
$\max_i\|\hat\eta_k^{(-i)}-\hat\eta_k\|_{P_0}=O_P(n^{-1/2})$ holds.
This provides a sufficient regular plug-in regime; a full general
verification theorem is not pursued here.
Under these conditions, a one-point-contamination expansion via the
chain rule for Hadamard derivatives
\citep[cf.][Chapter~20]{van_der_vaart1998asymptotic}
gives the uniform LOO linearization.
This regime covers fixed-weight ensembles and smooth sieves;
establishing Condition~(c) for complex data-adaptive procedures
requires further case-specific analysis.
\end{sremark}

\subsection*{S1.2\; Data-adaptive learners}

\begin{sremark}[Data-adaptive case]
For learners that are not Hadamard differentiable, Condition~(c) of
Proposition~\ref{sprop:hadamard} requires alternative verification.
Under a uniform bounded-differences stability condition
$\sup_x|\hat\eta_k(\mathcal{D}_n;x)-\hat\eta_k(\mathcal{D}_n^{(-i)};x)|
= O(n^{-1})$ a.s., the bilinear structure of $\Rrem$ gives
$\max_i|nb_i|=O(r_n)\to 0$.
This stability route corresponds most naturally to regimes where
LOO perturbations are negligible --- consistent with the strong-decay
case ($c_R=0$) rather than a regime of non-negligible remainder variance.
In the near-boundary regime ($c_R>0$), the moment conditions in the
LOO Remainder Linearization assumption require additional learner-specific
analysis beyond uniform stability bounds.
Assumption~5.1 is therefore stated as a direct condition on the LOO
perturbations to accommodate such case-by-case verification without
imposing a specific learner architecture.
\end{sremark}

\section{Expanded Proof of Theorem~5.2 (Bootstrap Variance Consistency)}
\label{sec:sm:boot}

\begin{stheorem}[Bootstrap Variance Consistency]
Under Assumptions~2.1--2.4 and Assumption~5.2 (Mallows-2) from the main
paper, the pairs bootstrap variance satisfies
\[
  \VarBoot(\Psihat) \xrightarrow{P} \mathrm{Var}(\Psihat-\Psitrue).
\]
Equivalently,
\[
  n\VarBoot(\Psihat) - \mathrm{Var}(T_n) \xrightarrow{P} 0,
  \qquad T_n=\sqrt{n}(\Psihat-\Psitrue).
\]
Thus the bootstrap targets the full sampling variance, including any
non-negligible remainder variance and covariance terms appearing in the
variance decomposition of Theorem~3.1 of the main paper.
\end{stheorem}

\begin{proof}
\textit{Step 1: Oracle bootstrap variance.}
Let $T_n = \sqrt{n}(\Psihat-\Psitrue)$ and $T_n^* = \sqrt{n}(\hat\Psi^*-\Psihat)$
be the centred bootstrap statistic.  Define the \emph{oracle} bootstrap variance
$V_n^{\rm oracle} := \mathrm{Var}^*[T_n^*\mid\mathcal{D}_n]$, the exact
conditional variance of $T_n^*$ given the observed data.

By the Mallows-2 bootstrap assumption, $d_2(T_n^*,T_n)\xrightarrow{P}0$, where
$d_2$ is the Wasserstein-2 distance.  Since Wasserstein-2 convergence is
equivalent to weak convergence plus convergence of second moments
\citep{bickel1981some}, and since $\EE^*[T_n^*\mid\mathcal{D}_n]\xrightarrow{P}
\EE[T_n]$ under the same Mallows-2 assumption,
\begin{equation}
  V_n^{\rm oracle}
  = \mathrm{Var}^*[T_n^*\mid\mathcal{D}_n]
  \;\xrightarrow{P}\; \mathrm{Var}(T_n).
  \label{eq:oracle_conv}
\end{equation}
The target $\mathrm{Var}(T_n)$ is exactly the variance of the full centred
estimator.  By Theorem~3.1 of the main paper,
$\mathrm{Var}(T_n)=\sigEIF + n\mathrm{Var}(\Rrem)+2n\mathrm{Cov}(\bar D_n,\Rrem)$,
so the bootstrap target automatically includes any non-negligible remainder
variance and covariance contributions.

\textit{Step 2: Oracle equals Monte Carlo in the $B\to\infty$ limit.}
The bootstrap variance estimator $\VarBoot(\Psihat)$ is defined as
the sample variance of $B$ conditionally i.i.d.\ bootstrap replicates
$\hat\Psi^{*1},\ldots,\hat\Psi^{*B}$:
\[
  \VarBoot(\Psihat)
  = \frac{1}{B-1}\sum_{b=1}^B
    \bigl(\hat\Psi^{*b}-\bar\Psi^*\bigr)^2,
  \quad
  \bar\Psi^* = \frac{1}{B}\sum_{b=1}^B\hat\Psi^{*b}.
\]
By the conditional law of large numbers applied to the bootstrap sample
variance around the bootstrap sample mean, as $B\to\infty$,
\[
  n\VarBoot(\Psihat)
  \;\xrightarrow{a.s.\mid\mathcal{D}_n}\;
  \mathrm{Var}^*(T_n^*\mid\mathcal{D}_n)
  = V_n^{\rm oracle}.
\]
Combining with~\eqref{eq:oracle_conv}:
\[
  n\VarBoot(\Psihat) \;\xrightarrow{P}\; \mathrm{Var}(T_n),
\]
equivalently
\[
  \VarBoot(\Psihat) \;\xrightarrow{P}\; \mathrm{Var}(\Psihat-\Psitrue).
\]
The bounded-second-moment condition in Assumption~5.2 supplies the
required uniform second-moment bound that justifies the exchange of limits
$B\to\infty$ and $n\to\infty$.
\end{proof}

\section{Expanded Proof of Theorem~5.4 (Jackknife Wald Interval Validity)}
\label{sec:sm:jk_ci}

\begin{stheorem}[Jackknife Wald Interval Validity]
Under the conditions of Theorem~5.1 (Jackknife Variance Consistency)
and Theorem~5.3 (Bootstrap Wald Interval Validity),
the studentized jackknife pivot
$T_n^{\rm JK} = (\Psihat - \Psitrue)/\sqrt{\VarJK} \Rightarrow N(0,1)$,
and the jackknife Wald interval is asymptotically valid.
\end{stheorem}

\begin{proof}
\textit{Step 1: Bootstrap pivot validity.}
By Theorem~5.3 and Assumption~5.3 (Near-Boundary CLT) from the main paper,
the bootstrap Wald pivot satisfies
$(\Psihat - \Psitrue)/\sqrt{\VarBoot} \Rightarrow N(0,1)$.

\textit{Step 2: Ratio convergence.}
By Theorem~5.1, $n\VarJK \xrightarrow{P} \sigEIF + c_R$.
By Theorem~5.2 and the negligible-cross-term condition of Theorem~5.3,
$n\VarBoot \xrightarrow{P} \sigEIF + c_R$.
Therefore $\VarJK/\VarBoot \xrightarrow{P} 1$.

\textit{Step 3: Transfer of validity.}
Write
\[
  T_n^{\rm JK}
  = \frac{\Psihat-\Psitrue}{\sqrt{\VarBoot}}
    \cdot \sqrt{\frac{\VarBoot}{\VarJK}}.
\]
The first factor converges in distribution to $N(0,1)$ (Step~1).
The second factor converges in probability to $1$ (Step~2).
Both quantities are measurable functions of the same data $\mathcal{D}_n$,
so Slutsky's theorem gives $T_n^{\rm JK}\Rightarrow N(0,1)$.

\textit{Strong-decay regime ($c_R = 0$).}
Both $n\VarJK$ and $n\VarBoot$ converge to $\sigEIF$;
the argument is identical.
\end{proof}

\section{Secondary Simulation: Strong-Decay Regime (AIPW)}
\label{sec:sm:aipw}

This section illustrates the strong-decay regime using a non-cross-fitted
augmented inverse probability weighted (AIPW) estimator in an i.i.d.\
observational study.  The purpose is to demonstrate that the variance ratio
$\hat\rho = \VarJK/\VarSand$ decreases to $1.0$ as $n$ grows when nuisance
estimation operates in the strong-decay regime
($n\cdot\mathrm{Var}(\Rrem)\to 0$), in contrast to the persistent
near-boundary behaviour of Table~1 in the main paper.

\subsection*{S4.1\; Data-generating process}

\begin{alignat*}{3}
  W &\sim N(0,1), \quad
  &A \mid W &\sim \mathrm{Bernoulli}(\expit(0.3W)),\\
  Y &= 0.5 + 0.4A + 0.3W + 0.15AW + \varepsilon, \quad
  &\varepsilon &\sim N(0, 0.5^2).
\end{alignat*}
True ATE: $\Psi_0 = 0.4$.  Nuisance models: outcome regression correctly
specified as $\mathrm{lm}(Y \sim A \times W)$; propensity enriched as
$\mathrm{glm}(A \sim W + W^2)$.  No cross-fitting ($V=1$) is used so that
the LOO and full-data fits are structurally comparable.  Under correctly
specified parametric nuisance at $O_P(n^{-1/2})$ convergence,
$n\cdot\mathrm{Var}(\Rrem)\to 0$, placing the estimator in the strong-decay
regime (Theorem~4.1 of the main paper).  By Cauchy--Schwarz,
$|n\,\mathrm{Cov}(\bar D_n,\Rrem)|
\le \{n\,\mathrm{Var}(\bar D_n)\}^{1/2}\{n\,\mathrm{Var}(\Rrem)\}^{1/2}
= O(1)\cdot o(1) = o(1)$,
so the cross-term condition of Remark~3.2 also holds here even without
cross-fitting.

\subsection*{S4.2\; Results}

\begin{table}[h]
\centering
\caption{AIPW secondary simulation: strong-decay regime.
  $\hat\rho = \bar{V}_{\rm JK}/\bar{V}_{\rm Sand}$ decreases monotonically
  to $1.0$, confirming $n\cdot\mathrm{Var}(\Rrem)\to 0$ under $O_P(n^{-1/2})$
  parametric nuisance.
  CP(Sand) and CP(JK): empirical 95\% CI coverage ($z_{\alpha/2}$ critical values).
  $R = 500$ replicates; MC-SE $\approx 0.010$.}
\label{tab:aipw}
\smallskip
\begin{tabular}{crrrrr}
\toprule
$n$ & Bias & MCSD & CP(Sand) & CP(JK) & $\hat\rho$ \\
\midrule
 200 & $+0.000$ & $0.070$ & $0.964$ & $0.968$ & $1.040$ \\
 500 & $-0.002$ & $0.045$ & $0.946$ & $0.946$ & $1.014$ \\
1000 & $-0.001$ & $0.031$ & $0.960$ & $0.960$ & $1.007$ \\
2000 & $-0.000$ & $0.023$ & $0.958$ & $0.958$ & $1.003$ \\
\bottomrule
\end{tabular}
\end{table}

$\hat\rho$ decreases monotonically from $1.040$ to $1.003$, consistent
with the sharp sandwich characterization (Theorem~4.1 of the main paper):
under $O_P(n^{-1/2})$ nuisance convergence, $n\VarSand \xrightarrow{P} \sigEIF$
and $n\VarJK \xrightarrow{P} \sigEIF$, so no correction beyond the sandwich
is needed.  All coverage probabilities lie within $\pm 2$ Monte Carlo
standard errors of $0.95$.  This contrasts with Table~1 of the main paper,
where $\hat\rho$ remains above $1.14$ across all $J$ under the SA-TMLE's
structural near-boundary remainder.

\section{Additional Proofs for Core Main-Text Results}
\label{sec:sm:core_proofs}

This section records short algebraic proofs for several results that are only
sketched in the main paper.  We write
\[
  \bar D_n := \frac{1}{n}\sum_{i=1}^n D^*(O_i;\eta^0),
  \qquad
  R_n := \Rrem,
\]
so that the von Mises expansion is
\[
  \Psihat - \Psitrue = \bar D_n + R_n.
\]
Throughout this section, $\mathbb E[D^*(O;\eta^0)] = 0$ and
$\mathrm{Var}\{D^*(O;\eta^0)\}=\sigEIF$.

\subsection*{S5.1\; Finite-sample variance decomposition (Theorem~3.1)}

\begin{proof}[Proof of Theorem~3.1 of the main paper]
By the von Mises expansion,
\[
  \mathrm{Var}(\Psihat-\Psitrue)
  = \mathrm{Var}(\bar D_n + R_n)
  = \mathrm{Var}(\bar D_n) + 2\,\mathrm{Cov}(\bar D_n,R_n) + \mathrm{Var}(R_n).
\]
Since the $D^*(O_i;\eta^0)$ are i.i.d., mean zero, and have variance
$\sigEIF$,
\[
  \mathrm{Var}(\bar D_n)
  = \frac{1}{n^2}\sum_{i=1}^n \mathrm{Var}\{D^*(O_i;\eta^0)\}
  = \frac{\sigEIF}{n}.
\]
Substituting yields
\[
  \mathrm{Var}(\Psihat-\Psitrue)
  = \frac{\sigEIF}{n} + 2\,\mathrm{Cov}(\bar D_n,R_n) + \mathrm{Var}(R_n),
\]
which is the claimed decomposition.
\end{proof}

\subsection*{S5.2\; Boundary variance gap and sandwich characterization}

\begin{proof}[Proof of Proposition~3.1 (Variance Gap at the Product-Rate Boundary)]
Assume the standard bilinear remainder structure together with the extra
conditions stated in the proposition:
\[
  n\,\mathrm{Cov}(\bar D_n,R_n) \to 0,
  \qquad
  c_R := \lim_{n\to\infty} n\,\mathrm{Var}(R_n) \in (0,\infty).
\]
Then the decomposition above gives
\[
  \mathrm{Var}(\Psihat-\Psitrue)
  = \frac{\sigEIF}{n} + o(n^{-1}) + \frac{c_R}{n} + o(n^{-1})
  = \frac{\sigEIF+c_R}{n} + o(n^{-1}).
\]
Hence
\[
  \mathrm{Var}(\Psihat-\Psitrue) - \frac{\sigEIF}{n}
  = \frac{c_R}{n} + o(n^{-1})
  = \Theta(n^{-1}),
\]
and division by $\sigEIF/n$ yields
\[
  \frac{\mathrm{Var}(\Psihat-\Psitrue)}{\sigEIF/n}
  = 1 + \frac{c_R}{\sigEIF} + o(1).
\]
Therefore the relative gap does not vanish unless $c_R=0$.
\end{proof}

\begin{proof}[Proof of Theorem~4.1 and Corollary~4.1]
The influence-function sandwich estimator is built from the empirical second
moment of the leading linear term, so under the bounded-moment assumptions in
Section~4 of the main paper,
\[
  n\,\VarSand(\Psihat) \xrightarrow{P} \sigEIF.
\]
Now consider the two regimes.

\smallskip
\noindent\emph{Strong decay.}
If $n\,\mathrm{Var}(R_n)\to 0$ and
$n\,\mathrm{Cov}(\bar D_n,R_n)\to 0$, then the decomposition in S5.1 gives
\[
  n\,\mathrm{Var}(\Psihat-\Psitrue)
  = \sigEIF + o(1),
\]
and the sandwich estimator is asymptotically correct for the full sampling
variance.

\smallskip
\noindent\emph{Near boundary.}
If $n\,\mathrm{Var}(R_n)\to c_R>0$ and
$n\,\mathrm{Cov}(\bar D_n,R_n)\to 0$, then
\[
  n\,\mathrm{Var}(\Psihat-\Psitrue)
  = \sigEIF + c_R + o(1),
\]
whereas still $n\,\VarSand(\Psihat) \xrightarrow{P} \sigEIF$.  Therefore
\[
  \mathrm{Var}(\Psihat-\Psitrue) - \VarSand(\Psihat)
  = \frac{c_R}{n} + o(n^{-1}),
\]
so the sandwich underestimates the total variance by a non-vanishing
first-order amount.

For the corollary, under the bilinear remainder structure of the main paper,
strictly faster than $n^{-1/4}$ convergence of each nuisance implies that the
remainder decays faster than $n^{-1/2}$, which is sufficient to force
$n\,\mathrm{Var}(R_n)\to 0$.  By contrast, the classical product-rate condition
$\|\hat\eta_1-\eta^0_1\|\,\|\hat\eta_2-\eta^0_2\|=o_P(n^{-1/2})$ controls
$\sqrt n R_n$ for centering but, without extra rate separation, does not by
itself imply the stronger variance requirement $n\,\mathrm{Var}(R_n)\to 0$.
\end{proof}

\subsection*{S5.3\; Full proof of the LOO expansion lemma (Lemma~5.1)}

\begin{proof}[Proof of Lemma~5.1 of the main paper]
Write $D_i:=D^*(O_i;\eta^0)$ and
\[
  \bar D_n^{(-i)} := \frac{1}{n-1}\sum_{j\neq i} D_j.
\]
The von Mises expansions for the full-sample and leave-one-out estimators are
\[
  \Psihat-\Psitrue = \bar D_n + R_n,
  \qquad
  \hat\Psi^{(-i)}-\Psitrue = \bar D_n^{(-i)} + R_n^{(-i)}.
\]
Subtracting,
\[
  \hat\Psi^{(-i)}-\Psihat = \bigl(\bar D_n^{(-i)}-\bar D_n\bigr) + \bigl(R_n^{(-i)}-R_n\bigr)
  = \bigl(\bar D_n^{(-i)}-\bar D_n\bigr) + b_i.
\]
Now
\[
  \bar D_n^{(-i)}
  = \frac{1}{n-1}\sum_{j\neq i} D_j
  = \frac{n\bar D_n-D_i}{n-1},
\]
so
\[
  \bar D_n^{(-i)}-\bar D_n
  = \frac{\bar D_n-D_i}{n-1}
  = -\frac{D_i-\bar D_n}{n-1}.
\]
Therefore
\[
  \hat\Psi^{(-i)}-\Psihat
  = -\frac{D_i-\bar D_n}{n-1} + b_i,
\]
which is the first displayed identity.

For the alternative representation, add and subtract $D_i/n$:
\[
  -\frac{D_i-\bar D_n}{n-1}
  = -\frac{D_i}{n} + \frac{n\bar D_n-D_i}{n(n-1)}
  = -\frac{D_i}{n} + a_i,
\]
where $a_i := (n\bar D_n-D_i)/\{n(n-1)\}$.
Hence
\[
  \hat\Psi^{(-i)}-\Psihat = -\frac{1}{n}D_i + b_i + a_i.
\]
It remains to verify the bounds on $a_i$.  Under Assumption~2.3
(Bounded Influence Function), $|D_i|\le M$ almost surely and therefore
$|\bar D_n|\le M$ almost surely.  Consequently
\[
  |a_i|
  \le \frac{n|\bar D_n|+|D_i|}{n(n-1)}
  \le \frac{2M}{n-1},
\]
uniformly in $i$, so $\max_i|a_i|=O_P(n^{-1})$.  Squaring and averaging,
\[
  \frac{1}{n}\sum_{i=1}^n a_i^2
  \le \frac{4M^2}{(n-1)^2}
  = O(n^{-2}).
\]
This proves the lemma.
\end{proof}

\subsection*{S5.4\; Full proof of the stabilization proposition (Proposition~5.1)}

\begin{proof}[Proof of Proposition~5.1 of the main paper]
Let
\[
  r_{n,i} := nb_i + \psi_{n,i},
  \qquad
  \bar r_n := n^{-1}\sum_{i=1}^n r_{n,i},
  \qquad
  \bar\psi_n := n^{-1}\sum_{i=1}^n \psi_{n,i}.
\]
Assumption~5.1 gives $\max_i|r_{n,i}|\to_P 0$ and
$\bar\psi_n=o_P(1)$.  Since $|\bar r_n|\le \max_i|r_{n,i}|$, also
$\bar r_n=o_P(1)$.  Rearranging the definition of $r_{n,i}$,
\[
  n(b_i-\bar b)
  = -\bigl(\psi_{n,i}-\bar\psi_n\bigr) + \bigl(r_{n,i}-\bar r_n\bigr).
\]
Hence
\[
  C_n
  := (n-1)\sum_{i=1}^n (b_i-\bar b)^2
  = \frac{n-1}{n^2}\sum_{i=1}^n
    \Bigl\{\bigl(\psi_{n,i}-\bar\psi_n\bigr)-\bigl(r_{n,i}-\bar r_n\bigr)\Bigr\}^2.
\]
Define the comparison quadratic form
\[
  Q_n := \frac{n-1}{n^2}\sum_{i=1}^n (\psi_{n,i}-\bar\psi_n)^2.
\]
By the elementary inequality $|u^2-v^2|\le 2|v||u-v|+|u-v|^2$,
\[
  |C_n-Q_n|
  \le \frac{n-1}{n^2}\sum_{i=1}^n
    \Bigl[2|\psi_{n,i}-\bar\psi_n|\,|r_{n,i}-\bar r_n| + (r_{n,i}-\bar r_n)^2\Bigr].
\]
Since $|r_{n,i}-\bar r_n|\le 2\max_j|r_{n,j}|$,
\[
  |C_n-Q_n|
  \le \frac{2(n-1)}{n^2}\max_j|r_{n,j}|\sum_{i=1}^n |\psi_{n,i}-\bar\psi_n|
      + \frac{4(n-1)}{n}\max_j r_{n,j}^2.
\]
Also
\[
  \frac{1}{n}\sum_{i=1}^n |\psi_{n,i}-\bar\psi_n|
  \le \Bigl(\frac{1}{n}\sum_{i=1}^n (\psi_{n,i}-\bar\psi_n)^2\Bigr)^{1/2}
  \le \Bigl(\frac{1}{n}\sum_{i=1}^n \psi_{n,i}^2\Bigr)^{1/2} + |\bar\psi_n|
  = O_P(1)
\]
by Assumption~5.1(ii).  Therefore the first term is $o_P(1)$ because
$\max_j|r_{n,j}|=o_P(1)$, and the second term is also $o_P(1)$.  Hence
$C_n - Q_n \xrightarrow{P} 0$.

Next observe that
\[
  Q_n
  = \frac{n-1}{n}\left\{\frac{1}{n}\sum_{i=1}^n \psi_{n,i}^2 - \bar\psi_n^2\right\}.
\]
By Assumption~5.1(i)--(ii), $\bar\psi_n^2=o_P(1)$ and
$n^{-1}\sum_i \psi_{n,i}^2\to_P c_R$, so $Q_n \xrightarrow{P} c_R$.
Combining gives $C_n\to_P c_R$, as claimed.
\end{proof}

\subsection*{S5.5\; Full proof of the jackknife consistency theorem (Theorem~5.1)}

\begin{proof}[Proof of Theorem~5.1 of the main paper]
Write $D_i:=D^*(O_i;\eta^0)$, $x_i:=D_i-\bar D_n$, and $y_i:=b_i-\bar b$.
By Lemma~5.1,
\[
  \hat\Psi^{(-i)}-\Psihat = -\frac{x_i}{n-1} + b_i.
\]
Averaging over $i$ and using $\sum_i x_i=0$ gives
$\bar\Psi_{(-\cdot)}-\Psihat = \bar b$.  Therefore
\[
  \hat\Psi^{(-i)}-\bar\Psi_{(-\cdot)} = -\frac{x_i}{n-1} + y_i.
\]
Substituting into the jackknife definition,
\[
  n\VarJK(\Psihat)
  = (n-1)\sum_{i=1}^n \left(-\frac{x_i}{n-1}+y_i\right)^2
  = A_n + B_n + C_n,
\]
where
\[
  A_n := \frac{1}{n-1}\sum_{i=1}^n x_i^2,
  \qquad
  B_n := -2\sum_{i=1}^n x_i y_i,
  \qquad
  C_n := (n-1)\sum_{i=1}^n y_i^2.
\]

\smallskip
\noindent\emph{Step 1: the influence-function term.}
Since $x_i=D_i-\bar D_n$,
\[
  A_n = \frac{n}{n-1}\left(\frac{1}{n}\sum_{i=1}^n D_i^2-\bar D_n^2\right)
  \xrightarrow{P} \sigEIF
\]
by the law of large numbers and $\mathbb E[D_i]=0$.

\smallskip
\noindent\emph{Step 2: the mixed term in the strong-decay regime.}
By Cauchy--Schwarz,
\[
  |B_n| \le 2\Bigl(\sum_{i=1}^n x_i^2\Bigr)^{1/2}
             \Bigl(\sum_{i=1}^n y_i^2\Bigr)^{1/2}.
\]
Now $\sum_i x_i^2 = O_P(n)$, while
$\sum_i y_i^2 \le \sum_i b_i^2 = o_P(n^{-1})$ under the strong-decay hypothesis.
Hence $B_n=o_P(1)$.  The same bound gives
$C_n \le (n-1)\sum_i b_i^2 = o_P(1)$.
Therefore, in the strong-decay regime,
$n\VarJK(\Psihat) = A_n + o_P(1) \xrightarrow{P} \sigEIF$.

\smallskip
\noindent\emph{Step 3: the mixed term in the near-boundary regime.}
Under Assumption~5.1,
\[
  y_i = -\frac{\psi_{n,i}-\bar\psi_n}{n} + \frac{r_{n,i}-\bar r_n}{n},
\]
with $\max_i|r_{n,i}|\to_P 0$.  Therefore
\[
  B_n
  = \frac{2}{n}\sum_{i=1}^n x_i(\psi_{n,i}-\bar\psi_n)
    - \frac{2}{n}\sum_{i=1}^n x_i(r_{n,i}-\bar r_n).
\]
Because $\sum_i x_i=0$, the $\bar\psi_n$ and $\bar r_n$ terms vanish, so
\[
  B_n = \frac{2}{n}\sum_{i=1}^n x_i\psi_{n,i}
        - \frac{2}{n}\sum_{i=1}^n x_i r_{n,i}.
\]
For the first sum, using Assumption~5.1(i)--(iii),
\[
  \frac{1}{n}\sum_{i=1}^n x_i\psi_{n,i}
  = \frac{1}{n}\sum_{i=1}^n D_i\psi_{n,i} - \bar D_n\bar\psi_n
  = o_P(1).
\]
Note the essential role of Assumption~5.1(iii) (cross-orthogonality)
here: without it, the first term on the right would not vanish and the
jackknife would fail to isolate $c_R$.  For the second sum,
\[
  \left|\frac{1}{n}\sum_{i=1}^n x_i r_{n,i}\right|
  \le \max_i|r_{n,i}|\cdot \frac{1}{n}\sum_{i=1}^n |x_i|
  = o_P(1),
\]
because $n^{-1}\sum_i |x_i| = O_P(1)$ and $\max_i|r_{n,i}|=o_P(1)$.
Thus $B_n=o_P(1)$ in the near-boundary regime as well.

Finally, Proposition~5.1 (proved in S5.4) gives
$C_n \xrightarrow{P} c_R$.
Combining with Step~1 and $B_n=o_P(1)$:
\[
  n\VarJK(\Psihat) = A_n + B_n + C_n \xrightarrow{P} \sigEIF + c_R.
\]
This proves both parts of the theorem.
\end{proof}

\section{Proof of Proposition~6.1 (ICC Amplification)}
\label{sec:sm:icc}

We provide the full derivation for the SA-TMLE data-generating process with
balanced clusters ($n_j = m$) and a normal random effect
$b_j \sim N(0,\sigma_b^2)$ independent of an individual-level error
$\varepsilon_{ijt}\sim N(0,\sigma_\varepsilon^2)$.  Throughout this section
$\rho_{\mathrm{ICC}} := \sigma_b^2/(\sigma_b^2+\sigma_\varepsilon^2)$.

Write the bilinear remainder as
\[
  \Rtwo = \frac{1}{J}\sum_{j=1}^J \tilde{R}_j,
  \quad
  \tilde{R}_j = \sum_{i,t}
    \underbrace{\bigl(\QYhat(S_{ijt},a,W_{ijt},t) - \QY(\cdot)\bigr)}_{\alpha_{ijt}}
    \cdot
    \underbrace{\Bigl(\tfrac{\gDhat(\cdot)}{g^0_\Delta(\cdot)} - 1\Bigr)}_{\beta_{ijt}}.
\]
Since clusters are i.i.d., $\mathrm{Cov}(\tilde{R}_j, \tilde{R}_{j'}) = 0$
for $j \neq j'$, so
\begin{equation}
  \mathrm{Var}(\Rtwo)
  = \frac{1}{J^2} \sum_{j=1}^J \mathrm{Var}(\tilde{R}_j)
  = \frac{1}{J}\,\mathrm{Var}(\tilde{R}_1).
  \label{eq:var_Rtwo}
\end{equation}

\paragraph{Within-cluster variance decomposition.}
Expand $\mathrm{Var}(\tilde{R}_j)$ by writing
$\tilde{R}_j = \sum_{(i,t)} \alpha_{ijt}\beta_{ijt}$.  For fixed nuisance
estimates and under the random-effects model,
$\alpha_{ijt} = f(S_{ijt}, W_{ijt}, t; b_j, \varepsilon_{ijt})$ where
the shared $b_j$ induces within-cluster dependence across $(i,t)$ pairs.
Specifically, write
\[
  \alpha_{ijt}\beta_{ijt}
  = \mu_{ijt} + \gamma_{ijt}\,b_j + \zeta_{ijt},
\]
where $\mu_{ijt} = \EE[\alpha_{ijt}\beta_{ijt} \mid b_j = 0]$,
$\gamma_{ijt}$ is the first-order sensitivity to $b_j$, and $\zeta_{ijt}$
is mean-zero given $b_j$.  Then
\[
  \tilde{R}_j
  = \underbrace{\sum_{i,t}\mu_{ijt}}_{\bar\mu}
    + b_j\underbrace{\sum_{i,t}\gamma_{ijt}}_{\bar\gamma}
    + \underbrace{\sum_{i,t}\zeta_{ijt}}_{\bar\zeta_j},
\]
with $b_j$ and $\bar\zeta_j$ independent.  Therefore
\[
  \mathrm{Var}(\tilde{R}_j)
  = \sigma_b^2\,\bar\gamma^2 + \mathrm{Var}(\bar\zeta_j).
\]

\paragraph{Structure of $\mathrm{Var}(\bar\zeta_j)$.}
Within a cluster, the terms $\zeta_{ijt}$ are independent given $b_j$ since
$\varepsilon_{ijt}$ are i.i.d.\ across $(i,t)$ conditionally on $b_j$.  Thus
\[
  \mathrm{Var}(\bar\zeta_j)
  = \sum_{i,t}\mathrm{Var}(\zeta_{ijt})
  = m\cdot T\cdot v_0
\]
for a per-observation residual variance $v_0 > 0$.

\paragraph{Structure of $\bar\gamma^2$.}
The summands $\gamma_{ijt}$ are equal to a common value $\gamma_0$ modulo
$O(n^{-1/2})$ terms (by the product-rate regularity of the nuisance fits),
so $\bar\gamma = mT\gamma_0 + O_P(m n^{-1/2})$.  Therefore
$\sigma_b^2\,\bar\gamma^2 = \sigma_b^2(mT)^2\gamma_0^2 + O(m^2 n^{-1})$.

\paragraph{Combining.}
Substituting back into~\eqref{eq:var_Rtwo},
\[
  \mathrm{Var}(\Rtwo)
  = \frac{mT\,v_0}{J} + \frac{\sigma_b^2(mT)^2\gamma_0^2}{J} + O(J^{-2}),
\]
where the $O(J^{-2})$ term absorbs the correction from the $O_P(m^2 n^{-1})$
nuisance approximation error.  Define
\begin{equation}
  V_0 \;:=\; \frac{mT\,v_0}{J},
  \qquad
  V_{\mathrm{ICC}} \;:=\;
  \frac{(\sigma_b^2+\sigma_\varepsilon^2)\,mT^2\gamma_0^2}{J}.
  \label{eq:vdef}
\end{equation}
With this definition,
\[
  \rho_{\mathrm{ICC}}\cdot m\cdot V_{\mathrm{ICC}}
  = \frac{\sigma_b^2}{\sigma_b^2+\sigma_\varepsilon^2}
    \cdot m \cdot \frac{(\sigma_b^2+\sigma_\varepsilon^2)\,mT^2\gamma_0^2}{J}
  = \frac{\sigma_b^2(mT)^2\gamma_0^2}{J},
\]
so the main-paper formula
\[
  \mathrm{Var}(\Rtwo)
  = V_0 + \rho_{\mathrm{ICC}}\cdot m \cdot V_{\mathrm{ICC}} + O(J^{-2})
\]
holds exactly.  Since $v_0 > 0$ and $\gamma_0^2(\sigma_b^2+\sigma_\varepsilon^2) > 0$,
both $V_0$ and $V_{\mathrm{ICC}}$ are strictly positive, and the gap
$\mathrm{Var}(\Rtwo) - V_0 = \rho_{\mathrm{ICC}}\cdot m\cdot V_{\mathrm{ICC}}
+ O(J^{-2})$ is strictly increasing in both $\rho_{\mathrm{ICC}}$ and $m$,
establishing the claimed ICC amplification.

\printbibliography